# Evolution of ferromagnetic and spin-wave resonances with crystalline order in thin films of full-Heusler alloy Co$_2$MnSi

Himanshu Pandey,[1] P. C. Joshi,[1] R. P. Pant,[2] R. Prasad,[1] S. Auluck,[1] and R. C. Budhani[1,2,a]
[1]*Department of Physics, Condensed Matter-Low Dimensional Systems Laboratory, Indian Institute of Technology, Kanpur-208016, India*
[2]*National Physical Laboratory, Dr. K. S. Krishnan Marg, New Delhi-110012, India*

We report the evolution of magnetic moment as well as magnetic anisotropy with crystalline order in Co$_2$MnSi thin films grown on (100) MgO by pulsed laser deposition. The films become more ordered as the annealing temperature ($T_A$) increases from 400 to 600 °C. The extent of $L2_1$ ordering in the films annealed at 600 °C is ≈96%. The static magnetization measurements by vibrating sample magnetometry shows a maximum moment of 4.95 $\mu_B$ per formula unit with low coercivity ($H_C \approx 65$ Oe) in the films annealed at 600 °C. A rigorous analysis of the azimuthal and polar angle dependent ferromagnetic resonance (FMR) measured at several temperatures allows determination of various anisotropy fields relevant to our system as a function of $T_A$. Finally, we have evaluated the exchange stiffness constant down to 100 K using spin wave modes in FMR spectra. We have also estimated the exchange energies as well as stiffness constant by *ab initio* calculations using the Korringa-Kohn-Rostoker method.

## I. INTRODUCTION

The Co-based Heusler alloys, which possess high magnetic moment and also high Curie temperature ($T_C$) along with theoretically predicted half metallic character, are potentially important for spintronics devices. However, due to atomic and/or antisite disorder, the values for spin polarization in epitaxial films of these alloys are found to be smaller than that obtained from various band structure calculations. Understanding the effect of structural disorder on the extent of spin splitting of the conduction band of Heusler alloys remains a challenging problem. Various groups have already reported the successful deposition of thin films of full-Heusler alloys, such as Co$_2$MnZ (Z: Si, Sn, Ge) on Al$_2$O$_3$,[1] Co$_2$FeSi on MgO (Ref. 2) and SrTiO$_3$,[3] Co$_2$MnSi on MgO (Ref. 4) and Al$_2$O$_3$.[5] The full-Heusler alloy Co$_2$MnSi (CMS) has attracted much attention because of its large energy bandgap of 0.4 eV for minority-spin band[6] and a high $T_C \approx 985$ K.[7] The effect of annealing temperature ($T_A$) on magnetic properties of CMS thin films and bulk samples has been studied very well but the obtained value of moment is still smaller than the theoretically predicted moment of 5.00 $\mu_B$ per formula unit (f.u.) for bulk $L2_1$ ordered CMS.[8] The measurements by Raphael *et al.*[9,10] on single crystals, arc melted polycrystals and $L2_1$ ordered polycrystalline thin films of CMS on glass show a moment which is the same as the theoretical value, in spite of a large (≈14%) Mn and Co site disorder in the latter two catagories of the samples. The site disorder also decreases the residual resistivity ratio from 6.5 (for single crystal) to a value of 1.4 for thin films. Several groups have used CMS as electrodes for magnetic tunnel junctions and obtained a spin polarization of 61% at 10 K,[11] and 89% at 2 K.[12] By using point contact Andreev reflection spectroscopy the spin polarization was found to be 55% at 4.2 K for CMS.[13] However these values of spin polarization are still smaller than the theoretically predicted value.[8,14]

Other than spin polarization, the properties, such as magnetic anisotropy, magnetic exchange interactions, and damping, must also be characterized to make use of these Heusler alloys as memory devices with high frequency response. A powerful, non-destructive technique which allows the measurement of the amplitude, dynamics, and anisotropy of magnetization vector **M** and individual spins, is microwave spectroscopy. Through the measurement of ferromagnetic resonance (FMR), one can get information about magnetic anisotropy, magnetic inhomogeneties, domain sizes, and magnetization relaxation processes. The FMR spectroscopy has been used to study different Heusler alloys, such as Co$_2$MnGe,[15] Co$_2$MnGa,[16] Co$_2$FeSi,[17] and manganites like La$_{0.7}$Sr$_{0.3}$MnO$_3$.[18] The magnetic properties like the anisotropy constants and damping constant were studied by FMR spectroscopy of CMS thin films on SiO$_2$ (Ref. 19) and MgO (Ref. 20) substrates. A quantitative understanding of the magneto-dynamic properties of the magnetic metals or alloys can be made by density functional theory (DFT) calculations. The *ab initio* calculations on CMS have been done extensively by several groups to understand the magnetic excitations, spin-wave stiffness, and variation of the Curie temperature. However, the effect of strain produced by substrate on the various physical properties still needs thorough investigations.

In this paper, we present the measurements of static magnetization and angular dependence of FMR to establish magnetic anisotropy in CMS films deposited on (100) MgO by pulsed laser ablation at 200 °C. We have optimized the growth of pure phase of CMS with structural and magnetic properties which are nearly equal to those of bulk by subsequently annealing the films at various temperatures. The increase in the post-deposition annealing temperature ($T_A$) increases the magnetic moment as well as $L2_1$ ordering parameter. The FMR approach also permits us to observe spin-wave resonance (SWR) spectra which are direct evidence of low energy excitations. By assuming the asymmetrical pinning state, we calculate the temperature dependent spin-wave stiffness constant of CMS films down to 100 K for the first time. The spin-wave stiffness constant, which illustrates the strength of the various exchange interactions within the sample, is an important quantity from the application point of view. The aim of this paper is to examine the static and dynamic magnetic properties of the CMS thin films. The structure of this paper is as follows: Sec. I, which covers the Introduction, has been presented already. Section II deals with details of the sample preparation and measurement procedures. All the results and their interpretation are given in Sec. III, which is further divided into subsections. The basic theory of FMR and how the FMR response can be used to calculate anisotropy and spin-wave stiffness is presented in Sec. III A. The X-ray diffraction and dc magnetization measurements are summarized in Sec. III B, and the experimental results of FMR are presented in Sec. III C. Finally, the *ab initio* calculation of exchange interactions and spin-wave stiffness constant by using the Korringa-Kohn-Rostoker (KKR) method are given in Sec. III D. In Sec. IV, we presented the conclusions drawn from these measurements and theoretical calculations.

[a]Electronic addresses: rcb@iitk.ac.in and rcb@nplindia.org.

## II. EXPERIMENTAL DETAILS

The epitaxial CMS thin films of 40 nm and 70 nm thickness were grown on single crystal (100) MgO substrate by pulsed laser ablation (laser fluency ≈ 3 J cm$^{-2}$) of a stoichiometric target of CMS made by arc melting. These films were deposited in an all-metal-seal high vacuum chamber equipped with a Ti-getter pump, and an ultrahigh vacuum compatible substrate heater which could heat the samples to ≈850 °C. We used a growth temperature of 200 °C followed by 1 h annealing at $T_A$ = 400, 500, and 600 °C in high vacuum to enhance crystallization and ordering in the films. In subsequent discussion, these samples will be identified as S(t/T), where t and T refer to thickness in nm and the annealing temperature in °C, respectively.

The crystallographic structure of the films was characterized using a PANalytical X'Pert PRO X-ray diffractometer equipped with a CuK$_{\alpha 1}$ source in $\theta$-$2\theta$ and $\varphi$ modes. For static magnetization measurements at room temperature with both in-plane ($H^{\parallel}$) and out-of-plane field ($H^{\perp}$) orientations, we have used a vibrating sample magnetometer (VSM) with a maximum field of 1.7 T. The FMR measurements were performed down to 100 K with a fixed microwave frequency (≈9.75 GHz), in a Bruker A 300 EPR spectrometer. The films for these measurements were mounted in the microwave cavity such that the applied static magnetic field could be aligned in the ($H^{\parallel}$) and ($H^{\perp}$) geometries by rotating the sample rod with a stepper motor. A diode detector measures the absorbed microwave power (P) by monitoring the reflected power from the cavity. In a standard FMR measurement, the external field is swept to make the magnetic sample go through the resonance condition, at which the microwave losses in the sample reach the peak. A modulation of amplitude 6 Oe and frequency 100 kHz was imposed on the static field to allow lock-in detection of the resonance condition. The recorded signal, which is proportional to the field derivative of the absorbed microwave power, is measured as a function of applied field.

## III. RESULTS AND DISCUSSION

### A. Anisotropy constants and spin wave modes

The dynamics of magnetization vector **M** under the influence of the microwave and static magnetizing field **H** can be described with the help of Fig. 1. We have used a spherical polar geometry where $\theta_M$ and $\theta_H$ are the polar angles of **M** and **H**, respectively, while the corresponding azimuthal angles are $\varphi_M$ and $\varphi_H$, respectively. The angle $\varphi_u$ defines the angle that in-plane anisotropy axis **H**$_u$ makes with the *x*-axis.

The time derivative of the magnetization around its equilibrium position is described by the Landau-Lifshitz-Gilbert equation as[21]

$$\frac{d\mathbf{M}}{dt} = -\gamma[\mathbf{M} \times (\mathbf{H} + \mathbf{H}_{eff})] + \frac{\alpha}{|\mathbf{M}|}\left(\mathbf{M} \times \frac{d\mathbf{M}}{dt}\right). \quad (1)$$

Here $\alpha$ is the Gilbert damping constant and $\gamma = g\mu_B/\hbar$, the gyromagnetic ratio with $g$ the spectroscopic splitting factor.

The total magnetic energy density $F$ of the film can be expressed as,[22]

$$\begin{aligned}F = &- M_s H[\sin\theta_M \sin\theta_H \cos(\varphi_M - \varphi_H) + \cos\theta_M \cos\theta_H] \\ &- (2\pi M_S^2 - K_\perp)\sin^2\theta_M - K_u \cos^2(\varphi_M - \varphi_u)\sin^2\theta_M \\ &- \frac{1}{8}[3 + \cos 4(\varphi_M - \varphi_4)]K_4 \sin^4\theta_M,\end{aligned} \quad (2)$$

where the first term is Zeeman energy, the second term the demagnetizing contribution, and the remaining two are the anisotropy energies. Here $M_S$ is the saturation magnetization, whereas $K_u$, $K_4$, and $K_\perp$ are the in-plane uniaxial, the fourfold and perpendicular anisotropy constants, respectively, with corresponding anisotropy fields $H_u = 2K_u/M_S$, $H_4 = 4K_4/M_S$, and $H_\perp = 2K_\perp/M_S$, respectively. The resonance frequency $\omega_r$ of the uniform precession mode obtained from the energy density is given by,

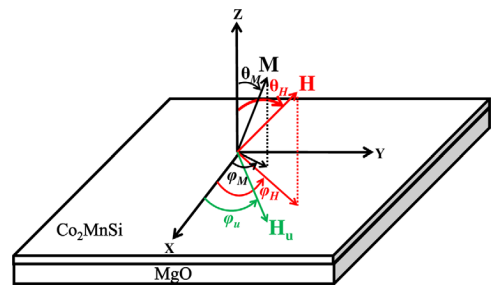

FIG. 1. (Color online) Schematic diagram of relevant vectors in spherical polar coordinates. The azimuthal angles $\varphi_M$, $\varphi_H$ are the in-plane angle between **M**, **H** and the *x*-axis, respectively. The angles $\theta_M$, $\theta_H$ are the polar angles between **M**, **H** and the *z*-axis, respectively. The $\varphi_u$ is the angle that in-plane anisotropy axis **H**$_u$ makes with the *x*-axis.

$$\omega_r = \frac{\gamma}{M \sin\theta_M} \left[ \frac{\partial^2 F}{\partial \theta_M^2} \frac{\partial^2 F}{\partial \varphi_M^2} - \left( \frac{\partial^2 F}{\partial \theta_M \partial \varphi_M} \right)^2 \right]^{1/2}. \quad (3)$$

Here all the derivatives are evaluated at the equilibrium positions of **M** and **H**. When the applied field is in the plane of the film, the resonance frequencies for uniform mode are given by,

$$\left(\frac{\omega_{res}^{\parallel}}{\gamma}\right)^2 = [H^{\parallel} \cos(\varphi_H - \varphi_M) + 4\pi M_{eff}$$
$$+ \frac{K_4}{2M_S}[3 + \cos 4(\varphi_M - \varphi_4)] + \frac{2K_u}{M_S}\cos^2(\varphi_M - \varphi_u)]$$
$$\times [H^{\parallel} \cos(\varphi_H - \varphi_M) + \frac{2K_4}{M_S}\cos 4(\varphi_M - \varphi_4)$$
$$+ \frac{2K_u}{M_S}\cos 2(\varphi_M - \varphi_u)], \quad (4)$$

where $4\pi M_{eff} = 4\pi M_S - 2K_\perp/M_S$ is the effective magnetization. For the out-of-plane applied magnetic field, the resonance frequencies are given by,

$$\left(\frac{\omega_{res}^{\perp}}{\gamma}\right)^2 = [H^\perp \cos(\theta_M - \theta_H) - 4\pi M_{eff} \cos 2\theta_M]$$
$$\times [H^\perp \cos(\theta_M - \theta_H) - 4\pi M_{eff} \cos^2 \theta_M]. \quad (5)$$

The FMR technique also allows us to study the spin-wave pattern formed by spin waves within the film. For a thin film of thickness $d$, the frequencies of standing waves in the case of out-of-plane applied magnetic field are given by

$$\left(\frac{\omega}{\gamma}\right) = H_n - 4\pi M_{eff} + \frac{D}{\gamma \hbar}\left(\frac{n\pi}{d}\right)^2. \quad (6)$$

Here $H_n$ is the resonance field for the spin wave mode of order $n$ and $D$ is the spin-wave stiffness constant. The integer $n$ corresponds to the number of half wavelengths of spin wave spectra within the thickness of the film. In the absence of SWR modes (i.e., $n=0$), Eq. (6) gives a uniform FMR mode. The spin-wave stiffness constant $D$ can be calculated by considering two different spin-wave modes in Eq. (6).

### B. Structural and magnetic measurements

First of all, we have characterized the crystallographic structure of our CMS films with X-ray diffraction. The crystal structure of the ordered full-Heusler alloy consists of four fcc sublattices with Co atoms at (1/4, 1/4, 1/4) and (3/4, 3/4, 3/4), the Mn atom at (0, 0, 0), and the Si atom at (1/2, 1/2, 1/2) in Wyckoff coordinates. Figure 2(a) shows the $\theta$-$2\theta$ diffraction profile of S(70/600), where the (200) and (400) reflections of the cubic phase of CMS are clearly seen. The intensities of these peaks increase with $T_A$ indicating a better crystallinity for the films annealed at higher temperature. The X-ray peaks for the $L2_1$ structure (space group $Fm\bar{3}m$) should have two types of superlattice reflection, i.e., (111) and (200) in addition to (220) fundamental reflection. The intensity ratio $I_{111}/I_{220}$ provides the measure of the $L2_1$ atomic site ordering in the Heusler alloys. The $\varphi$ scans for the (111) and (220) super-lattices of the sample S(70/600) are shown in the upper and lower panel of the Fig. 2(b), respectively. This gives the direct evidence of crystalline growth of S(70/600) in $L2_1$ structure with $\approx$96% ordering. The $\varphi$ values for the (111) and (220) planes of the CMS film are shifted by 45° with respect to those of the MgO planes. A similar epitaxiality with $L2_1$ ordering is also observed in S(40/600). By reducing the annealing temperature, the extent of ordering decreases, and the films annealed below 400 °C are amorphous.

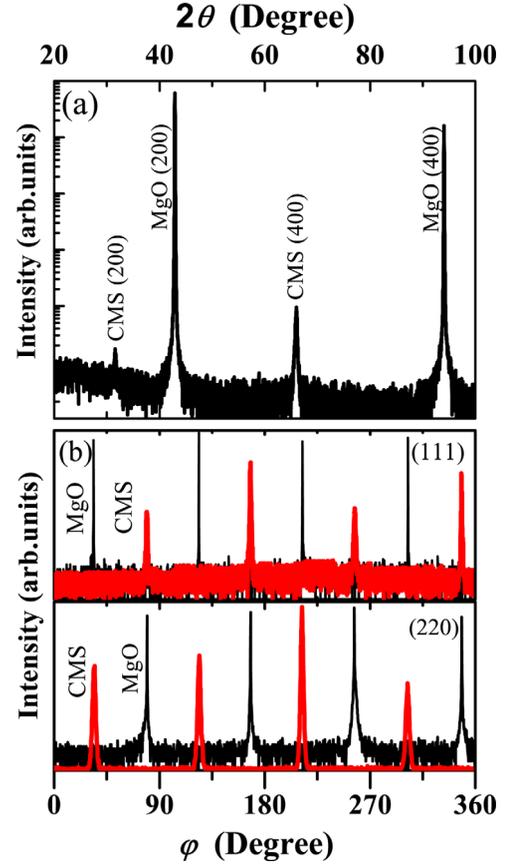

FIG. 2. (Color online) (a) X-ray diffraction $\theta$-$2\theta$ scan of the sample S(70/600). The (200) and (400) peaks of the CMS film and substrate MgO are marked in the diffraction pattern, (b) $\varphi$ scans for the (111) and (220) planes of CMS (red) and MgO (black) plotted in the upper and lower panels, respectively. The peaks are marked adjacent to their positions.

The hysteresis loops of 70 nm films with different $T_A$ measured at room temperature with in-plane magnetic fields are shown in Fig. 3. The magnetic moment of the CMS film increases on increasing the $T_A$ due to increase in the $L2_1$ ordering parameter. The inset of Fig. 3 shows the hysteresis loop of S(70/600) for both $H^{\parallel}$ and $H^{\perp}$ geometries, which clearly indicates it to be a soft magnet with in-plane easy axis and coercivity of $\approx$65 Oe. The magnetic moment per f.u. for S(40/600) and S(70/600) is 4.78 $\mu_B$ and 4.95 $\mu_B$, respectively. These values are in reasonable agreement with the predicted value of 5.00 $\mu_B$/f.u. according to the Slater-Pauling rule.[8]

### C. Dynamic magnetic measurements

We can determine the values of anisotropy constants very precisely from the angular variation of the resonance fields by using Eqs. (4) and (5). The left-hand side of Fig. 4 shows the room temperature FMR spectra of the 70 nm film annealed at different temperatures for $H^{\parallel}$ configuration while the data for $H^{\perp}$ are shown on the right-hand side. The observation of the sharp resonance peaks with narrow linewidth indicates excellent magnetic homogeneity of the samples, except for S(70/400) where spurious peaks appear at lower field for in-plane geometry. A noteworthy feature of $H^{\perp}$ data is the presence of well resolved SWR modes, which can be seen in the 40-fold magnified out-of-plane data of S(70/600) [Fig. 4(a)].

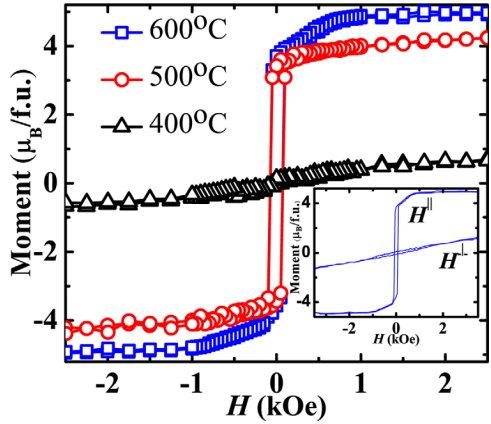

FIG. 3. (Color online) The room temperature magnetic hysteresis loops of 70 nm CMS films with different $T_A$ for in-plane magnetic field. The inset shows the hysteresis loops for S(70/600) for $H^{\parallel}$ and $H^{\perp}$ orientations.

Figures 5(a) and 5(b) show the azimuthal and polar angular dependence of the resonance field for S(70/600), respectively. Figure 5(a) shows a fourfold symmetry with easy axis along the [110] direction. An excellent fit to Eq. (4) of these data is found with $\varphi_u$ and $\varphi_4$ equal to 45° for the films annealed at 500 and 600 °C, whereas for S(70/400) we found $\varphi_u = 17°$ and $\varphi_4 = 45°$. This is in agreement with the X-ray measurements which show that the (100) plane of the CMS film is rotated by 45° with respect to the (100) plane of MgO. The fitting also gives the value of the in-plane anisotropy constants $K_u = 4.52 \times 10^4$ erg/cm$^3$ and $K_4 = 1.14 \times 10^5$ erg/cm$^3$, respectively, for S(70/600). These numbers are comparable with the earlier result obtained from Brillouin light scattering spectroscopy[23] and FMR technique[20] for the thin films of CMS. The in-plane anisotropy fields for all the films are summarized in the inset of Fig. 5(b). We note that the resonance fields are a strong function of $T_A$ and hence are related to the crystallite size, the extent of ordering, magnetostriction, and the stress due to epitaxy. Thus, the increase in the $T_A$ leads to an increase in the anisotropy constants, since the annealing improves the crystallinity and ordering of the $L2_1$ structure. However, in all the films the easy axis remains along the [110] direction except 400 °C films.

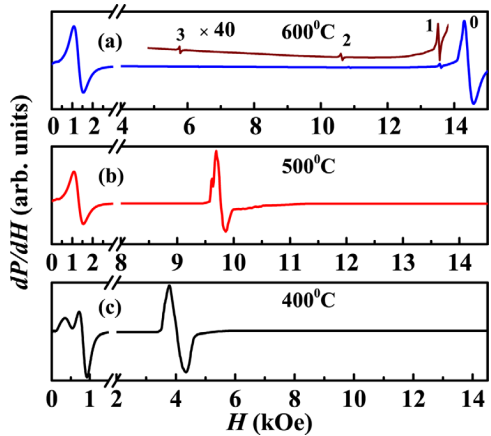

FIG. 4. (Color online) FMR spectra of 70 nm CMS films annealed at 600, 500, and 400 °C are shown in panels (a), (b), and (c), respectively. Each panel is divided into two parts; left and right correspond to in-plane and out-of-plane orientation of the magnetic field, respectively. Panel (a) also shows SWR modes with the corresponding spin-wave number after 40-fold magnification of the data.

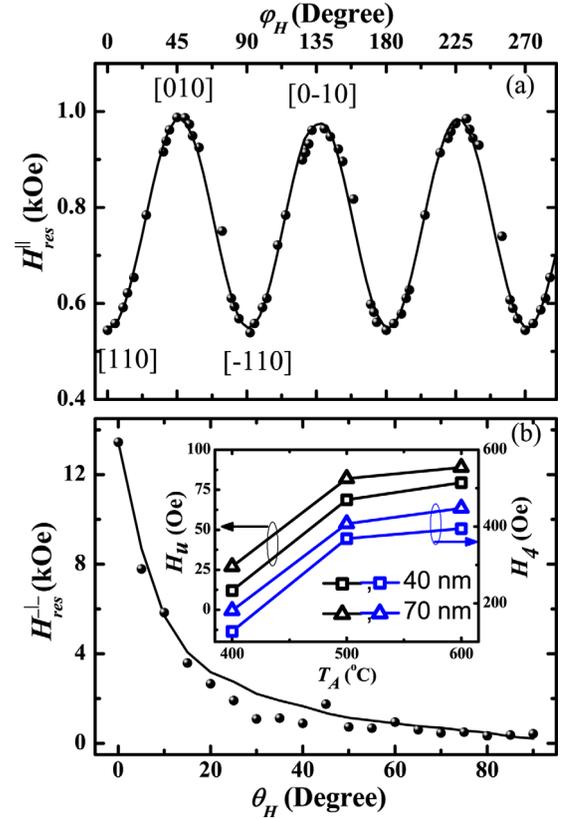

FIG. 5. (Color online) The azimuthal (a) and polar (b) angular dependence of the resonance field for S(70/600). The solid lines are the fits to Eq. (4) for panel (a) and Eq. (5) for panel (b). The inset in panel (b) shows the anisotropy fields $H_u$ and $H_4$ as a function of $T_A$. The arrows show the corresponding anisotropy fields.

The polar angular variation of FMR [see Fig. 5(b)] clearly shows the hard axis of magnetization to be along the [001] direction. A very strong field is needed to rotate **M** from the easy axis to the sample's normal direction. The polar variation of the resonance field for S(70/600) is in good agreement with Eq. (5) with fitting parameters $g = 2.11$ and $4\pi M_{eff} = 14.5$ kOe. The nearly same $g$ value is obtained for S(40/600) but with a slightly lesser value of $4\pi M_{eff}$ (=13.7 kOe). Using the saturation magnetization determined from hysteresis loops in Fig. 3, we can calculate the perpendicular anisotropy field by the relation $4\pi M_{eff} = 4\pi M_S - 2K_{\perp}/M_S$. We observe the $K^{\perp}$ to be negative for all the films, implying an in-plane easy axis. The anisotropy fields ($H_{\perp}$) are 1260 Oe for S(40/600) and 1610 Oe for S(70/600). These values are a bit smaller than the previously reported[20] value for Cr-buffered CMS films on MgO due to the excess tensile strain on CMS by MgO in our case. The variation of FMR linewidhts ($\Delta H_{PP}$) for azimuthal and polar angular configuration for S(70/600) are shown in the inset of Fig. 6(a). The $\Delta H_{PP}$ values remain nearly constant for azimuthal angular variation, whereas for the case of polar angular variation it increases on decreasing the temperature. One important feature of the FMR spectra of thin magnetic films is the observation of satellite peaks on the lower field side of the main resonance similar to the peaks seen in Fig. 4(a). The origin of these peaks has been attributed to a nonuniform field seen by the spins as a function of film thickness.[24] These peaks can be indexed by only odd or only even or both even and odd numbers. We have determined the conditions from SWR spectrum by using relative intensities of asymmetric pinning suggested by Puszkarski.[25] In our work, we have assigned both even and odd values to the spin-wave mode number $n$ and obtained a linear dependence of absorption fields $H_n$ with $n^2$ [see Fig. 6(a)]. This also confirms the asymmetrical pinning conditions at two different film surfaces. We also notice a considerably narrower linewidth of SWR peaks as compared to the linewidth of the FMR peak. This suggests a local variation in strain, magnetorestriction or small local irregularities in the magnetic lattice, or a very small variation in the film thickness across the plane of the film. The spin-wave stiffness constant $D$ at room temperature extracted from Fig. 6(a), is 4.87 meV nm$^2$ and 5.22 meV nm$^2$ for S(40/600) and S(70/600), respectively. We have fitted this temperature dependent $D$ using the Dyson equation expressed as follows:[26]

$$D(T) = D(0)(1 - CT^{5/2}). \quad (7)$$

Here $D(0)$ is the stiffness constant value at $T=0$ K and $C$ is the constant. The fitting of $D(T)$ in Fig. 6(b) shows the $T^{5/2}$ dependence with $D(0) = (4.98 \pm 0.04)$ meV nm$^2$ and $C = (1.47 \pm 0.08) \times 10^{-8}$ K$^{-5/2}$ for S(40/600), whereas $D(0) = (5.34 \pm 0.11)$ meV nm$^2$ and $C = (1.42 \pm 0.03) \times 10^{-8}$ K$^{-5/2}$ for S(70/600). The obtained values of $D$ are comparable with the value $\approx 5.75$ meV nm$^2$ reported by Hamrle et al.[27] although we have not used any buffer layer in the fabrication of our films. The $D(0)$ values for 40 nm and 70 nm films as a function of $T_A$ are shown in the inset of Fig. 6(b). With increasing $T_A$ the films become more ordered which increases the magnetic moment and the exchange interaction between the sublattices, which further increases $D$.

### D. Computational details and estimation of stiffness constant

The *ab initio* calculations on CMS with cubic crystal structure (space group $Fm\bar{3}m$), were carried out by using the Korringa-Kohn-Rostoker Green's function method within the spin-polarized scalar-relativistic Hamiltonian.[28] A $22 \times 22 \times 22$ mesh (834 k-points) and 30 energy points on the complex energy path were used to obtain self-consistent ground-state potential with local spin density approximation. The interactions within a sphere of radius $3a$ ($a$ is the lattice parameter of CMS) were considered for the calculation of exchange energies. The spin wave energy $E(\mathbf{q})$ is related to the exchange interaction as given below:

$$E(q) = \frac{4\mu_B}{m} \sum_{j \neq 0} J_{oj}[1 - e^{i\mathbf{q}\cdot\mathbf{R}_{oj}}], \quad (8)$$

where $m$ is the magnitude of the local magnetic moment per atom and $J_{ij}$ is the exchange energy between $i$th and $j$th sites with $\mathbf{R}_{0j} = \mathbf{R}_0 - \mathbf{R}_j$, the translation vector between site 0 and $j$ and $\mathbf{q}$, a vector in the corresponding Brillouin zone. For the cubic systems with small $\mathbf{q} \to 0$, we have $E(\mathbf{q}) \approx Dq^2$, where $q = |\mathbf{q}|$. Thus the spin-wave stiffness $D$ is given by the curvature of the spin-wave dispersion $E(q)$ at $q=0$. We have calculated the spin-wave stiffness constant for CMS using the expression[29]

$$D = \frac{2}{3} \sum_j |R_{oj}|^2 \frac{J_{0j}}{\sqrt{m_i m_j}} e^{(-\eta R_{0j}/a)} \quad (9)$$

in the limit $\eta \to 0$; where $\eta$ is the damping parameter.

Figure 7(a) shows the variation of exchange energies of CMS as a function of distance ($R_{ij}/a$) with $a = 0.5655$ nm, the bulk $L2_1$ ordered CMS lattice parameter. The exchange interaction is calculated between Co1-Co1, Co1-Co2, Mn-Co, and Mn-Mn sublattices, where Co1 and Co2 are the two types of Co atoms at different Wyckoff positions. The strongest interaction is found between the Mn-Co sublattices followed by the interaction between Co1-Co2 sublattices. The coupling between the nearest neighbor sites has the maximum contribution while all other interactions contribute negligibly. The strength of Co1-Co1 and Mn-Mn interactions are very small and nearly same in magnitude. Hence, the stability of ferromagnetism in CMS is mainly governed by Mn-Co interactions between the Mn atom and the eight nearest Co atoms. The interactions between Co-Co and Mn-Mn sublattices are of different signs for different distances between atoms due to Ruderman-Kittel-Kausya-Yoshida oscillations as shown in Fig. 7(a). We have calculated the spin-wave stiffness constant for CMS by using Eq. (9) for different pairs of sublattices on varying $\eta$ values between 0.5 and 1.5 and then the corresponding $D$ at $\eta = 0$ is estimated by taking the quadratic fit [see inset of Fig. 7(a)]. By summing the $D(\eta=0)$ values for above mentioned pairs, the value of $D$ for CMS is found to be 5.23 meV nm$^2$, which is comparable to that obtained by Thoene et al.[30] We have also investigated the effect of lattice parameter ($a$) of film on the magnetic properties of CMS, since the thin films grown MgO will have a larger value of $a$ due to in-plane tensile strain. The calculated magnetic moments of individual sites as well as for total CMS unit cell are shown in Fig. 7(b). There is a significant increase in the moments of Mn sites which increases with $a$, whereas on the other hand the moments for the Co site decreases as the value of $a$ increases above the bulk value. These two site-specific moments (Co and Mn) balance each other so that the net magnetic moment remains unchanged. The total moment changes slightly with lattice constant but remains almost constant near a value of 5 $\mu_B$ up to $\pm 6\%$ change in the lattice parameter. Figure 7(c) shows the variation of $D$ with lattice parameter. The value of $D$ remains constant for the values of lattice constant larger than bulk because the system attains the full magnetization but for lower lattice constants it increases rapidly due to a stronger hybridization between the $d$ orbitals of Co and Mn atoms as well as the increase in exchange interactions. This clearly shows that our experimental values of moment and spin-wave stiffness constant are close to the theoretically estimated ones even after considering the effect of strain on CMS.

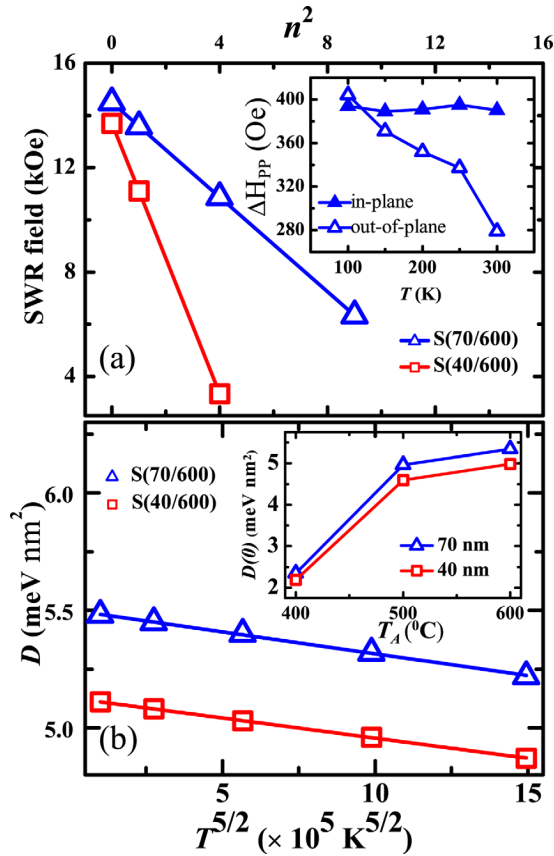

FIG. 6. (Color online) (a) Dependence of SWR absorption field on $n^2$, and (b) temperature dependence of the spin-wave stiffness coefficient ($D$) for S(40/600) and S(70/600). The inset in panel (a) shows the temperature dependence of azimuthal and polar angular FMR linewidths for S(70/600) and the inset in panel (b) shows the $T_A$ dependence of $D(0)$ values.

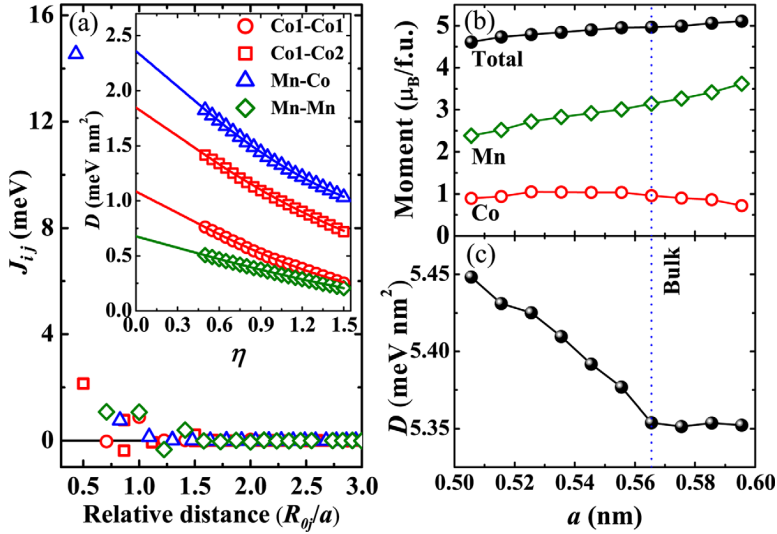

FIG. 7. (Color online) (a) The variation of interatomic exchange energies $J_{ij}$ for CMS with the distance between interacting sites. The inset shows the calculated spin-wave stiffness constant for different pairs of sublattices as a function of $\eta$. The lines are the quadratic fit to the data. The variation of (b) site-specific and total moment of CMS as well as (c) spin-wave stiffness constant with the lattice parameter. The blue dotted line indicates the lattice constant for bulk CMS.

## IV. SUMMARY

We have addressed the magnetic state of CMS thin films having a different degree of $L2_1$ ordering using static magnetization measurements and ferromagnetic resonance absorption in a microwave cavity. The magnetic moment ($\approx 4.95$ $\mu_B$/f.u.) of the well-ordered crystalline films is in agreement with the calculated value for this half-metallic ferromagnetic. Further, these films are magnetically soft with in-plane magnetic easy axis and coercive field $H_C \approx 45\text{-}150$ Oe. We have precisely determined the anisotropy parameters via in-plane and out-of-plane angular dependence of resonance fields. A good fit of all experimental data, using appropriate values of the magnetic coefficients which describe the free energy, is obtained. We found $K_\perp = -8.18 \times 10^5$ erg/cm$^3$, $K_u = 4.52 \times 10^4$ erg/cm$^3$, and $K_4 = 1.14 \times 10^5$ erg/cm$^3$ for S(70/600). The presence of even and odd spin-wave modes indicates that the state of pinning is asymmetric at the surface and interface of the films. The linear dependence of $H_n$ on $n^2$ gives the stiffness constant value $D(0)$ of $(5.34 \pm 0.11)$ meV nm$^2$ for S(70/600). We have determined, for the first time, the temperature dependence of $D$ for CMS, which follows a $T^{5/2}$ dependence. The exchange energies as well as the $D$ values have been obtained from first principle calculations. We have also calculated the effect of lattice parameter on the magnetic moment and $D$, which is consistent with our experimental results for CMS film with tensile strain.

## ACKNOWLEDGMENTS

This research has been supported by a grant from the Department of Information Technology (DIT), Government of India. H.P. acknowledges financial support from Indian Institute of Technology (IIT) Kanpur and the Council for Scientific and Industrial Research (CSIR), Government of India. H.P. would like to thank P. K. Rout for the fruitful discussion and M. Bahl for technical help. R.C.B.

acknowledges the J C Bose National Fellowship of the Department of Science and Technology. We are very grateful to Professor Hubert Ebert for allowing us the use of the SPRKKR code.

[1] U. Geiersbacha, A. Bergmanna, and K. Westerholt, J. Magn. Magn. Mater. **240**, 546 (2002).
[2] H. Schneider, G. Jakob, M. Kallmayer, H. J. Elmers, M. Cinchetti, B. Balke, S. Wurmehl, C. Felser, M. Aeschlimann, and H. Adrian, Phys. Rev. B **74**, 174426 (2006).
[3] Anupam, P. C. Joshi, P. K. Rout, Z. Hossain, and R. C. Budhani, J. Phys. D: Appl. Phys. **43**, 255002 (2010).
[4] Y. Sakuraba, T. Iwase, K. Saito, S. Mitani, and K. Takanashi, Appl. Phys. Lett. **94**, 012511 (2009).
[5] L. J. Singh, Z. H. Barber, Y. Miyoshi, Y. Bugoslavsky, W. R. Branford, and L. F. Cohen, Appl. Phys. Lett. **84**, 2367 (2004).
[6] S. Ishida, S. Fujii, S. Kashiwagi, and S. Asano, J. Phys. Soc. Jpn. **64**, 2152 (1995).
[7] P. J. Brown, K. U. Neumann, P. J. Webster, and K. R. A. Ziebeck, J. Phys.: Condens. Matter **12**, 1827 (2000).
[8] I. Galanakis, P. H. Dederichs, and N. Papanikolaou, Phys. Rev. B **66**, 174429 (2002).
[9] M. P. Raphael, B. Ravel, Q. Huang, M. A. Willard, S. F. Cheng, B. N. Das, R. M. Stroud, K. M. Bussmann, J. H. Claassen, and V. G. Harris, Phys. Rev. B **66**, 104429 (2002).
[10] M. P. Raphael, B. Ravel, M. A. Willard, S. F. Cheng, B. N. Das, R. M. Stroud, K. M. Bussmann, J. H. Claassen, and V. G. Harris, Appl. Phys. Lett. **79**, 4396 (2001).
[11] S. Kämmerer, A. Thomas, A. Hütten, and G. Reiss, Appl. Phys. Lett. **85**, 79 (2003).
[12] Y. Sakuraba, J. Nakata, M. Oogane, H. Kubota, Y. Ando, A. Sakuma, and T. Miyazaki, Jpn. J. Appl. Phys. **44**, L1100 (2005).
[13] L. J. Singh, Z. H. Barber, A. Kohn, A. K. Petford-Long, Y. Miyoshi, Y. Bugoslavsky, and L. F. Cohen, J. Appl. Phys. **99**, 013904 (2006).
[14] S. Picozzi, A. Continenza, and A. J. Freeman, Phys. Rev. B **66**, 094421 (2002).
[15] M. Belmeguenai, F. Zighem, Y. Roussigné, S.-M. Chérif, P. Moch, K. Westerholt, G. Woltersdorf, and G. Bayreuther, Phys. Rev. B **79**, 024419 (2009).
[16] C. Yu, M. J. Pechan, D. Carr, and C. J. Palmstrøm, J. Appl. Phys. **99**, 08J109 (2006).
[17] M. Oogane, R. Yilgin, M. Shinano, S. Yakata, Y. Sakuraba, Y. Ando, and T. Miyazaki, J. Appl. Phys. **101**, 09J501 (2007).
[18] M. Golosovsky, P. Monod, P. K. Muduli, and R. C. Budhani, Phys. Rev. B **76**, 184413 (2007).
[19] R. Yilgin, M. Oogane, Y. Ando, and T. Miyazaki, J. Magn. Magn. Mater. **310**, 2322 (2007).
[20] R. Yilgin, Y. Sakuraba, M. Oogane, S. Mizumaki, Y. Ando, and T. Miyazaki, Jpn. J. Appl. Phys. **46**, L205 (2007).
[21] O. Acher, S. Queste, M. Ledieu, K.-U. Barholz, and R. Mattheis, Phys. Rev. B **68**, 184414 (2003).
[22] M. Farle, Rep. Prog. Phys. **16**, 755 (1998).
[23] O. Gaier, J. Hamrle, S. J. Hermsdoerfer, H. Schultheiβ, B. Hillebrands, Y. Sakuraba, M. Oogane, and Y. Ando, J. Appl. Phys. **103**, 103910 (2008).
[24] C. Kittel, Phys. Rev. **110**, 1295 (1958).
[25] H. Puszkarski, Prog. Surf. Sci. **9**, 191 (1979).
[26] F. J. Dyson, Phys. Rev. **102**, 1217 (1956); **102**, 1230 (1956).
[27] J. Hamrle, O. Gaier, S.-G. Min, B. Hillebrands, Y. Sakuraba, and Y. Ando, J. Phys. D: Appl. Phys. **42**, 084005 (2009).
[28] H. Ebert, see http://olymp.cup.unimuenchen.de/ak/ebert/SPRKKR for a description of the 2005 Munich SPR-KKR package, Version 5.4.
[29] M. Pajda, J. Kudrnovsky, I. Turek, V. Drchal, and P. Bruno, Phys. Rev. B **64**, 174402 (2001).
[30] J. Thoene, S. Chadov, G. Fecher, C. Felser, and J. Kübler, J. Phys D: Appl. Phys. **42**, 084013 (2009).